\documentclass[twoside]{article}

\usepackage{koopman_format}
\usepackage{verbatim}

\usepackage{tikz-cd}

\usepackage[numbers]{natbib}
\usepackage{microtype}

\usepackage[hide=false,setmargin=true,marginparwidth=0.5in]{marginalia}

\usepackage{multicol}
\usepackage{}

\usepackage{graphicx} 
\usepackage{subfigure}

\usepackage{url}
\usepackage{bm}
\usepackage{framed}

\usepackage{amsmath}
\usepackage{amsfonts}
\DeclareMathAlphabet{\mathpzc}{OT1}{pzc}{m}{it}

\usepackage{stackrel}

\usepackage[shortlabels]{enumitem}

\usepackage{xcolor}
\usepackage{tikz}

\usepackage{rotating}

\usepackage{amsthm}
\usepackage{mathtools}

\usepackage{algorithm}
\usepackage{algpseudocode}

\usepackage[overlay]{textpos}

\theoremstyle{theorem}

\theoremstyle{lemma}

\theoremstyle{definition}

\theoremstyle{conjecture}

\usetikzlibrary{calc,trees,positioning,arrows,chains,shapes.geometric,%
    decorations.pathreplacing,decorations.pathmorphing,shapes,%
    matrix,shapes.symbols}

\tikzset{
>=stealth',
  punktchain/.style={
    rectangle, 
    rounded corners, 
    draw=black, very thick,
    text width=10em, 
    minimum height=3em, 
    text centered, 
    on chain},
  line/.style={draw, thick, <-},
  element/.style={
    tape,
    top color=white,
    bottom color=blue!50!black!60!,
    minimum width=8em,
    draw=blue!40!black!90, very thick,
    text width=10em, 
    minimum height=3.5em, 
    text centered, 
    on chain},
  every join/.style={->, thick,shorten >=1pt},
  decoration={brace},
  tuborg/.style={decorate},
  tubnode/.style={midway, right=2pt},
}

\usepackage{hyperref}

\usepackage{dblfloatfix}
\usepackage{float}
\usepackage{inconsolata}

\usepackage{hyperref}
\hypersetup{
  colorlinks=true,
  citecolor=blue,
  urlcolor=magenta,
}
\begin{document}

\twocolumn[
  \ourtitle{Extended Koopman Models}
  \ourauthor{Span Spanbauer \And Ian Hunter}
  \ouraddress{MIT \And MIT}

]

\begin{abstract}

We introduce two novel generalizations of the Koopman operator method of nonlinear dynamic modeling. Each of these generalizations leads to greatly improved predictive performance without sacrificing a unique trait of Koopman methods: the potential for fast, globally optimal control of nonlinear, nonconvex systems. The first generalization, \textit{Convex Koopman Models}, uses convex rather than linear dynamics in the lifted space. The second, \textit{Extended Koopman Models}, additionally introduces an invertible transformation of the control signal which contributes to the lifted convex dynamics. We describe a deep learning architecture for parameterizing these classes of models, and show experimentally that each significantly outperforms traditional Koopman models in trajectory prediction for two nonlinear, nonconvex dynamic systems.
\end{abstract}



\section{Introduction}
The Koopman operator approach to nonlinear dynamic modeling, introduced in 1931~\citep{koopman1931hamiltonian}, is a method of globally linearizing nonlinear dynamic systems. It has seen widespread application throughout science, particularly in analyzing fluid flows~\citep{mezic2013analysis,bagheri2013koopman,muld2012flow,schmid2011applications}, but also in neuroscience~\citep{brunton2016extracting}, algorithmic trading~\citep{mann2016dynamic}, and computer vision~\citep{kutz2015multi}. Prior to the introduction of techniques for estimating Koopman operators beginning in 2004~\citep{mezic2004comparison,schmid2008dynamic}, Koopman operator theory was primarily applied as a tool in stochastic systems theory, most notably in Von Neumann's mean ergodic theorem~\citep{neumann1932proof,arbabi2018introduction}.

Koopman operator methods have recently been applied to the optimal control of nonlinear systems~\citep{proctor2018generalizing} including fluids~\citep{peitz2019koopman} and electromechanical systems~\citep{korda2018linear}. Since Koopman models globally linearize nonlinear dynamics, globally optimal control trajectories can be found quickly via linear model predictive control using quadratic programming~\citep{muske1993model}.

However, these applications use finite dimensional approximations of the exact infinite dimensional Koopman operators. These approximate models often fail to predict trajectories over even moderate timescales. For example, we observe that for a simulated particle driven by a random force in a double-well potential, a model based on a 256 dimensional approximation of the Koopman operator predicts a trajectory which diverges from the ground truth in only about 10 timesteps. This rapid breakdown in predictive ability limits the applicability of model predictive control based off of Koopman methods.

We address this breakdown in predictive performance with two novel extensions to the Koopman operator approach.

\noindent {\bf Contributions.} This paper introduces two extensions to the Koopman operator method of nonlinear dynamic modeling, greatly improving predictive performance without sacrificing the potential for fast globally optimal control. Specifically, it presents the following contributions:
\vspace*{-5pt}
\begin{enumerate}
\item This paper introduces \textit{Convex Koopman Models}, which use convex rather than linear dynamics in the lifted space. This improves predictive performance while still enabling fast globally optimal trajectory optimization via convex optimization.

\item This paper introduces \textit{Extended Koopman Models}, which combine convex lifted dynamics with a contribution from an invertibly transformed control signal. This invertible transformation is learned from a class of functions parameterized by the lifted state, and generalizes the standard linear contribution of the control signal to the lifted dynamics in traditional Koopman models. This generalization increases predictive performance while still enabling fast globally optimal trajectory optimization; one optimizes the transformed control signal and then applies the inverse to obtain the optimal control signal.

\item This paper describes a class of deep learning architectures which act as high-capacity parameterizations of these models.

\item This paper provides experimental results showing that \textit{Convex Koopman Models} and \textit{Extended Koopman Models} each significantly outperform traditional Koopman models in trajectory prediction on two nonlinear, nonconvex problems.

\end{enumerate}

\newpage

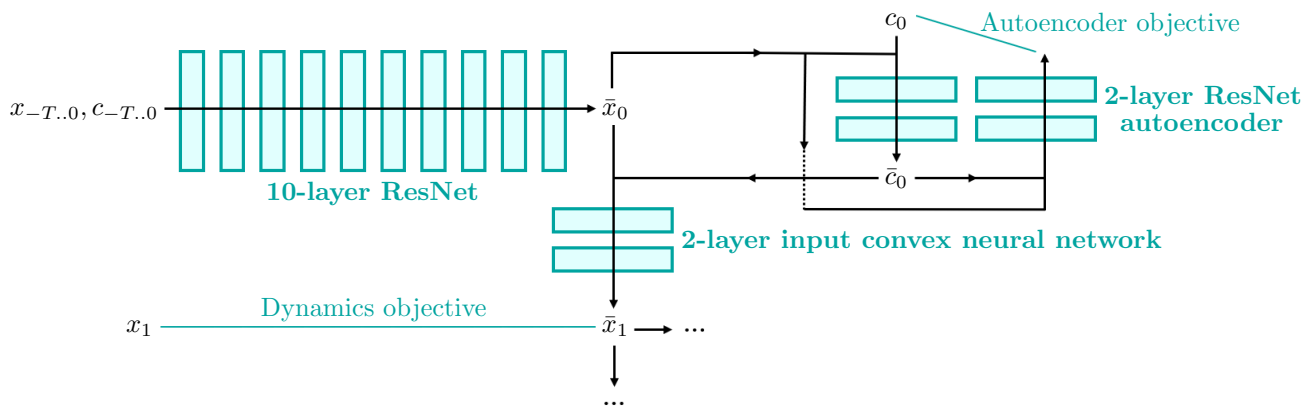
\begin{figure*}[!t]

\centering

\begin{picture}(468,175)
\put(-4,0){\includegraphics[width=170mm]{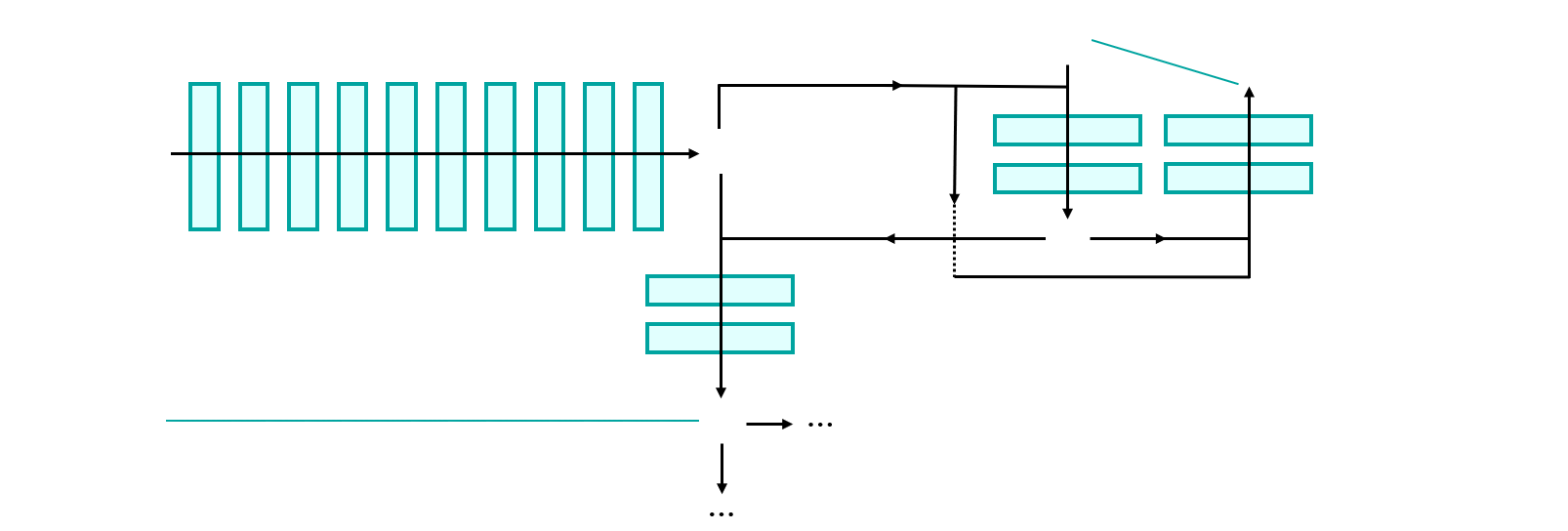}}
\put(95,165){\textbf{Example architecture for Extended Koopman modeling}}
\definecolor{chart_aqua}{RGB}{0, 164, 160}
\put(357,146){\textcolor{chart_aqua}{Autoencoder objective}}

\put(404,118){\textcolor{chart_aqua}{\textbf{2-layer ResNet}}}
\put(410,108){\textcolor{chart_aqua}{\textbf{autoencoder}}}

\put(244,63.5){\textcolor{chart_aqua}{\textbf{2-layer input convex neural network}}}

\put(87,82){\textcolor{chart_aqua}{\textbf{10-layer ResNet}}}

\put(85,38){\textcolor{chart_aqua}{Dynamics objective}}

\put(-10,114){$x_{-T..0},c_{-T..0}$}
\put(34,32){$x_1$}
\put(214,32){$\bar x_1$}
\put(214,114){$\bar x_0$}
\put(321,148){$c_0$}
\put(321,88){$\bar c_0$}

\end{picture}

\vspace{0mm}

\caption{This architecture was used for our experiments with \textit{Extended Koopman Models}. A short history of observations (T=2) is passed into a 10-layer deep residual network~\citep{he2016deep} implementing the lifting function $\phi$, yielding the lifted state $\bar x_0$. The control signal for the current timestep $c_0$ is passed through a 2-layer ResNet encoder along with $\bar x_0$ implementing the approximately invertible control transformation parameterized by $\bar x_0$, yielding the transformed control signal $\bar c_0$. This is then passed through a 2-layer ResNet decoder along with $\bar x_0$ and compared with $c_0$ to generate an objective function enforcing approximate invertibility. The lifted state $\bar x_0$ and control $\bar c_0$ are passed through a 2-layer input convex neural network~\citep{amos2017input} implementing the convex dynamics in the lifted space, yielding the next timestep of the lifted state $\bar x_1$. This is compared with the true state to generate an objective function enforcing consistency of the dynamics.}
\label{fig:architecture}
\end{figure*}

\section{Background}
\label{sec:background}
The Koopman operator approach to nonlinear dynamic modeling proceeds by finding a nonlinear transformation which lifts the system state into a new space in which the dynamics are globally linear. This global linearization is highly attractive, since it enables the use of linear systems theory to understand and control the system. 

Consider system dynamics given in state-space form
\begin{align*}
\dot x(t) = f(x(t),u(t)).
\end{align*}
The Koopman operator approach involves finding a function $\phi$ mapping the state $x$ to the lifted state $\bar x$, and finding linear operators $A$ and $B$ which define linear dynamics on the lifted state
\begin{align*}
\bar x(t) &= \phi(x(t))\\
\dot {\bar x}(t) &= A \bar x(t) + B u(t)
\end{align*}
such that the dynamics defined by $f$ are consistent with the linear dynamics in the lifted space; that is, such that the following diagram commutes for all $t_1$ and $t_2$.

\begin{centering}

\tikzcdset{column sep/normal=4.5cm}
\tikzcdset{row sep/normal=1.5cm}
\begin{tikzcd}x(t_1)\arrow[r, "\text{nonlinear time evolution by }f"]\arrow[d, "\phi"]& x(t_2)\arrow[d, "\phi"] \\ \bar x(t_1) \arrow[r, "\text{linear time evolution by }A \text{ and } B"]&  \bar x(t_2) \end{tikzcd}

\end{centering}

Unfortunately, for general nonlinear dynamic systems this is only possible using an infinite dimensional lifted space. For intuition, consider the method of Carleman linearization~\citep{kowalski1991nonlinear}, which can be considered a Koopman operator method in which the lifting function maps the vector of state variables to the infinite dimensional vector of all products of integer powers of state variables. This enables the function $f$ defining the nonlinear dynamics to be expressed as a linear combination of the lifted state variables via Taylor expansion.

Thus in practice Koopman methods only achieve approximate global linearization; one lifts the state to a finite dimensional space in which linear dynamics are only approximately consistent with the true dynamics. Using a higher dimensional lifted space yields a better approximation but is more computationally expensive, ultimately limiting the time horizon over which the predictions of this method can be trusted. We will see in Section~\ref{sec:particle} that for even a very simple nonlinear, nonconvex system, the Koopman approach breaks down after about 10 timesteps even for a high capacity lifting function mapping into a large lifted space of dimension 256.

In the next section we consider extensions to the Koopman operator method which significantly increase its modeling capacity while retaining attractive properties. 

\section{Extended Koopman Models}
We aim to increase the representational power of Koopman-style models without sacrificing an important trait: the ability to perform fast, globally optimal trajectory optimization. To do this we generalize traditional Koopman models in two ways.

First, to obtain \textit{Convex Koopman Models} we replace the usual linear dynamics in the lifted space with more general convex dynamics. To perform trajectory optimization on this lifted convex dynamics we would solve a convex optimization problem rather than the quadratic programming problem which arises for linear dynamics. Since $n$-dimensional convex optimization problems can be solved to accuracy $\epsilon$ in $\mathcal{O}(n \: \text{log} \: 1/\epsilon)$ iterations via, for example, the center of gravity algorithm~\citep{bubeck2014convex}, we would still be able to perform fast global trajectory optimization.

Second, to obtain \textit{Extended Koopman Models} we additionally replace the usual linear control contribution to the lifted dynamics with a linear contribution from an invertibly transformed control input, where the invertible function is parameterized by the current lifted state. We bound the range of this transformation; this would enable us to enforce constraints on the control signal as bounds in the convex optimization problem in the transformed space. We may then use the inverse transformation to recover the optimal control signal.

Taken together, the dynamics of an \textit{Extended Koopman Model} in the lifted space are
\begin{align*}
\dot {\bar x}(t) = g(\bar x(t),h_{\bar x(t)}(u(t)))
\end{align*}
where $g$ is a convex function of its inputs and $h_{\bar x(t)}$ is invertible and bounded.

Having described our model classes, we now move to describe our parameterization and training methods.

\section{Architecture}
While there are many potential ways to parameterize \textit{Convex} and \textit{Extended Koopman Models}, we designed an architecture focusing on high capacity and trainability; a schematic of this architecture is shown in Fig.~\ref{fig:architecture}.

To implement the lifting function $\phi$, we used a 10-layer densely connected deep residual network~\citep{he2016deep}. The hidden layers have width 512, while the output of the last layer, the lifted state, has dimension 256.

For the convex dynamics $g$, we used a 2-layer densely connected input convex neural network~\citep{amos2017input} using hidden layers with output dimension 256. We pass the inputs to all previous layers as inputs to each subsequent layer in order to increase capacity and learnability. Learnability of the convex dynamics is particularly important since gradients will need to pass through many copies of this network during each training step: one copy for each of the 10-25 timesteps we train through.

To implement $h$ and $h^{-1}$, the approximately invertible control transformation parameterized by the lifted state, we used an autoencoder architecture consisting of a 2-layer densely connected residual network encoder and a 2-layer densely connected residual network decoder. Each hidden layer has width 512, and the latent space is of the same dimension as the control input. The inputs to the encoder are the control signal and the lifted state, and the inputs to the decoder are the transformed control signal and the lifted state.

For our experiments with \textit{Convex Koopman Models}, we used the same architecture except with no autoencoder; the control signal is passed directly into the input convex neural network implementing the convex dynamics.

For our experiments involving traditional Koopman models, we use the same architecture but replace the input convex neural network with a linear layer.

\section{Training}

For all models, we use a supervised contribution to the objective function enforcing consistency of the lifted dynamics. We require that the head of the lifted state match one or more dynamic quantities derived from the unlifted state on each of $n$ timesteps, where $n$ was gradually increased from 10 to 25 over the course of training; deviation is penalized with a $L_2$ loss.

For the \textit{Extended Koopman Models}, we furthermore use contributions to the objective function enforcing approximate invertibility and boundedness of the control transformation. At each training step we sample random valid control values $c_i$, random valid transformed control values $\bar c_i$, and random observed lifted states $\bar x_i(t)$. We then enforce cyclic consistency $c_i = h^{-1}_{\bar x_i(t)}(h_{\bar x_i(t)}(c_i))$ and $\bar c_i = h_{\bar x_i(t)}(h^{-1}_{\bar x_i(t)}(\bar c_i))$ with a $L_2$ loss. We enforce the boundedness of our transformations with a quadratic penalty for out-of-bounds signals. In our experiments we were able to enforce cyclic consistency and boundedness to about 1\% error without significantly impacting predictive capability.

Models were trained to convergence in PyTorch~\citep{NEURIPS2019_9015} with an Adam optimizer~\citep{kingma2014adam}, using an NVIDIA~RTX~2080~Ti GPU. For the \textit{Extended Koopman Models}, convergence took 2 hours for the simpler system described in Section~\ref{sec:particle}, and 24 hours for the more complex system described in Section~\ref{sec:quadbot}.

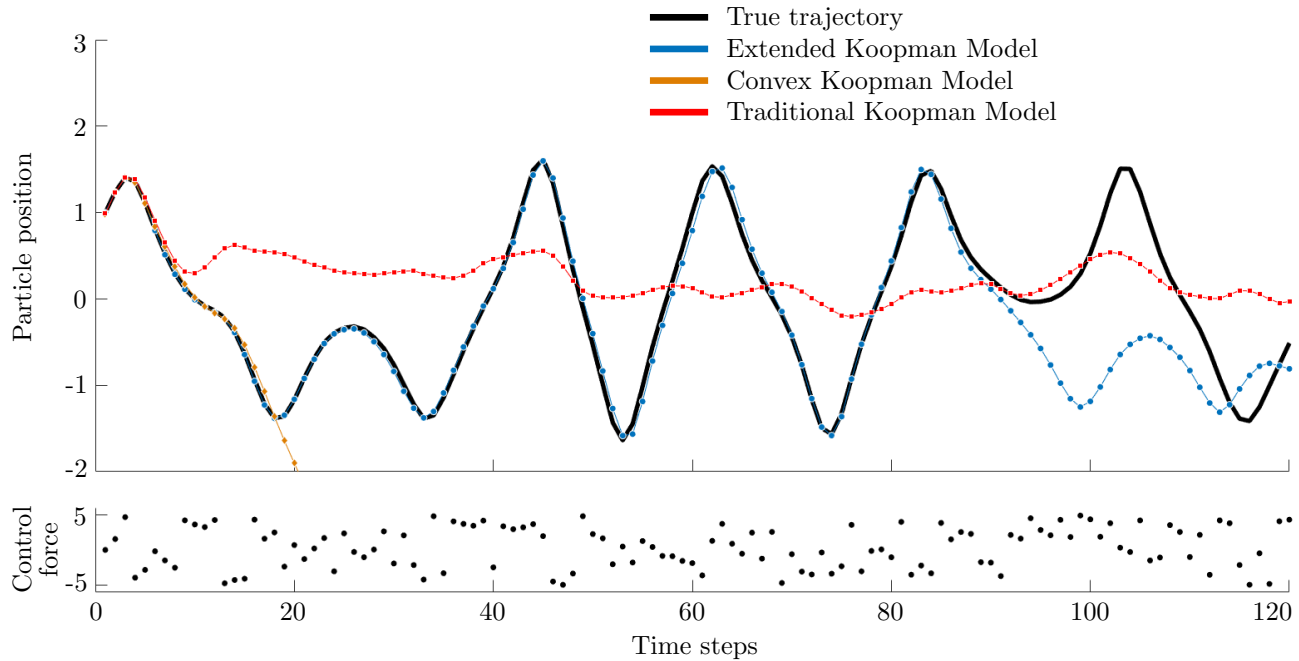
\begin{figure*}[!t]
\centering

\begin{picture}(422,205)
\put(0,0){\includegraphics[width=160mm]{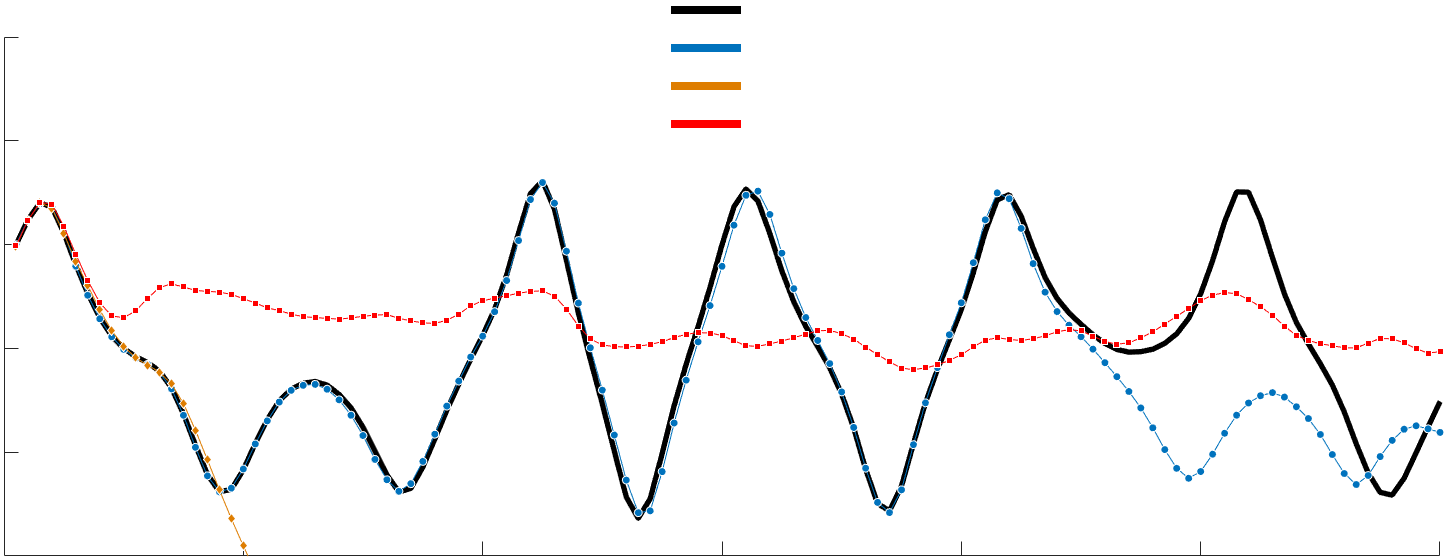}}
\put(55,195){\textbf{Trajectory prediction for a particle in a double-well potential}}
\put(240,170){True trajectory}
\put(240,158){Extended Koopman Model}
\put(240,146){Convex Koopman Model}
\put(240,134){Traditional Koopman Model}

\put(204,-68){Time steps}
\put(-30,50){\rotatebox{90}{Particle position}}

\put(-10,-1){-2}
\put(-10,31){-1}
\put(-7,63){0}
\put(-7,96){1}
\put(-7,128){2}
\put(-7,160){3}

\put(-1,-55){0}
\put(71,-55){20}
\put(146.5,-55){40}
\put(222,-55){60}
\put(297,-55){80}
\put(370,-55){100}
\put(439,-55){120}

\put(0,-45){\includegraphics[width=160mm]{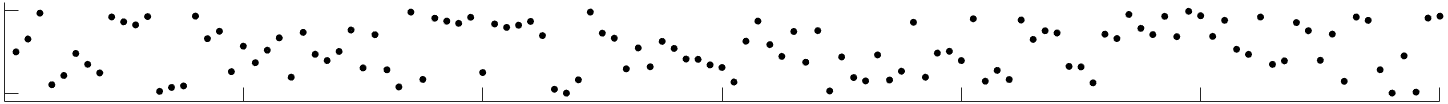}}

\put(-10,-44){-5}
\put(-7,-19){5}

\put(-30,-45){\rotatebox{90}{Control}}
\put(-20,-39){\rotatebox{90}{force}}
\end{picture}

\vspace{25mm}

\caption{Consider the problem of predicting the trajectory of a particle in a double-well potential given an initial state and a sequence of random forces on the particle. The potential minima are at -1 and 1, and there is a local maxima at 0; notice how the particle sometimes remains trapped within a well and sometimes travels over the middle potential barrier. A traditional Koopman model fails to predict the trajectory beyond about 10 time steps. A \textit{Convex Koopman Model} involving convex dynamics in the lifted space predicts the trajectory out to about 20 time steps. An \textit{Extended Koopman Model} involving both convex dynamics in the lifted space as well as an approximately invertible control transformation is capable of predicting the trajectory out to about 90 time steps. Notice that once the \textit{Extended Koopman Model} diverges from the true trajectory, it still predicts a qualitatively reasonable trajectory unlike the other two model types. The \textit{Extended Koopman Model} predicts that the particle will barely make it over the potential barrier at the 95\textsuperscript{th} time step, when it in fact barely does not.}
\label{fig:well-results}
\end{figure*}

\section{Experimental Details and Results}
\label{sec:results}
We fit a traditional Koopman model, a \textit{Convex Koopman Model}, and an \textit{Extended Koopman Model} to the dynamics of each of two simulated physical systems. 

The first system, a particle in a double-well potential acted on by an external force, is detailed in Section~\ref{sec:particle}. This system was chosen to display the performance of the various models in a simple nonlinear, nonconvex context. 

The second system, a quadrupedal robot with twelve actuators, is detailed in Section~\ref{sec:quadbot}. This system was chosen to demonstrate scalability to complex systems with many control inputs.

In both cases we will find that the \textit{Extended Koopman Model} performs remarkably well, and that the \textit{Convex Koopman Model} significantly outperforms the traditional Koopman model.

 \begin{figure*}[!t]
\centering

\vspace{-5.9cm}
\begin{picture}(422,400)
\put(0,150){\includegraphics[width=159.3mm]{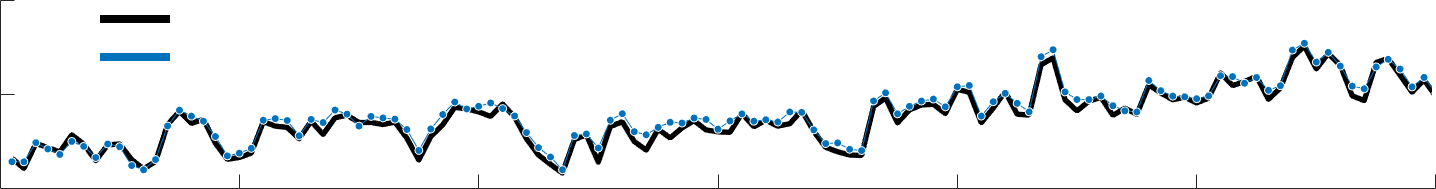}}
\put(0,75){\includegraphics[width=159.3mm]{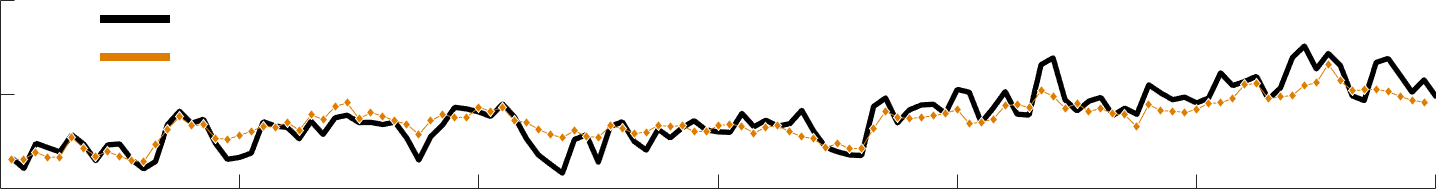}}
\put(0,0){\includegraphics[width=159.3mm]{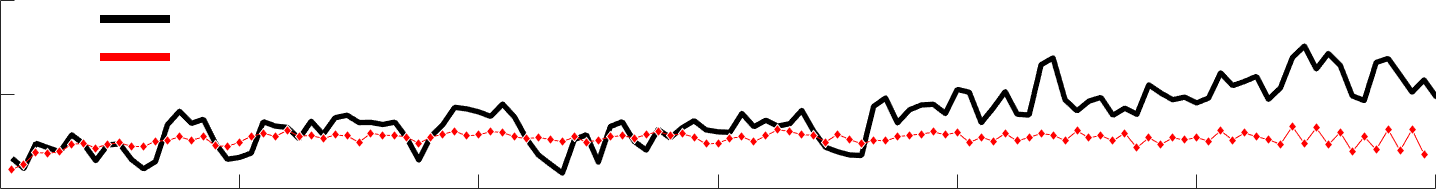}}

\put(98,223){\textbf{Trajectory prediction for a quadrupedal robot}}
\put(58,201){True trajectory}
\put(58,188){Extended Koopman Model}

\put(58,126){True trajectory}
\put(58,113){Convex Koopman Model}

\put(58,51){True trajectory}
\put(58,38){Traditional Koopman Model}


\put(204,-25){Time steps}
\put(-32,-5){\rotatebox{90}{Height of center of mass of quadrupedal robot (m)}}

\put(-15,-2){0.1}
\put(-15,56){0.3}

\put(-15,73){0.1}
\put(-15,131){0.3}

\put(-15,153){0.1}
\put(-15,206){0.3}

\put(-1,-11){0}
\put(71,-11){20}
\put(146.5,-11){40}
\put(222,-11){60}
\put(297,-11){80}
\put(370,-11){100}
\put(439,-11){120}

\end{picture}

\vspace{12mm}

\caption{Consider the problem of predicting the trajectory of the center of mass of a quadrupedal robot, given an initial state and a sequence of random control signals to the 12 actuators. Traditional Koopman models fail to predict the trajectory beyond a few time steps. A \textit{Convex Koopman Model} involving convex dynamics in the lifted space captures some of the dynamics and manages to follow the general trend of the trajectory, but fails to resolve fine details. An \textit{Extended Koopman Model} involving both convex dynamics in the lifted space as well as an approximately invertible control transformation is capable of predicting the trajectory out to 120 time steps with only small deviations.}
\label{fig:quadbot-results}
\end{figure*}
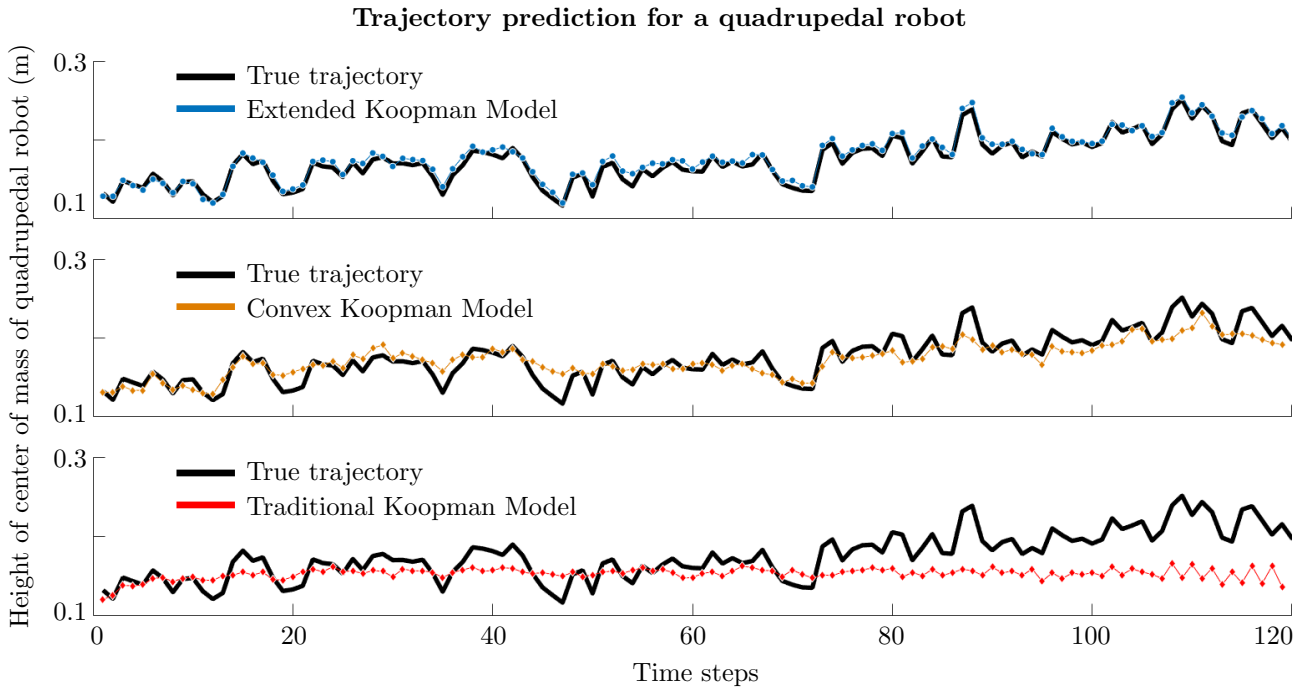

\subsection{Particle in a double-well potential}
\label{sec:particle}

We consider a simple, yet nonlinear and nonconvex system described by the following potential
\begin{align*}
U(x) = (1-x^2)^2
\end{align*}
which, when acted upon by a control force c, leads to the following dynamics:
\begin{align*}
\ddot x = 4 x (1 - x^2) + c .
\end{align*}
For this nondissipative system, driving it with a random control force will cause the energy to grow without bound. Therefore we add a dissipative term to the dynamics, yielding the following forward-Euler scheme which we repeat 10 times per timestep with $\delta t=0.1$
\begin{align*}
\dot x_{t+1/10} &= (\dot x_{t} + \frac{\delta t}{10} (4 x (1 - x^2) + c)) 0.99^{1/10}\\
x_{t+1/10} &= x_{t} + \frac{\delta t}{10} \: \dot x_{t+1/10} .
\end{align*}

We chose uniformly random control inputs and simulated the dynamics for 500,000 timesteps to use as training data. We then trained each model to convergence and compared their respective performance.

We observe, as illustrated in Fig.~\ref{fig:well-results}, that the \textit{Extended Koopman Model} is able to accurately capture this nonlinear, nonconvex dynamics, making good predictions out to about 90 timesteps. The \textit{Convex Koopman Model} predicts accurately out to 20 timesteps, a significant improvement over the traditional Koopman model, which predicts accurately out to about 10 timesteps.

 \begin{figure}[!t]
\centering
\begin{picture}(300,155.5)
\put(0,0){\includegraphics[width=83mm]{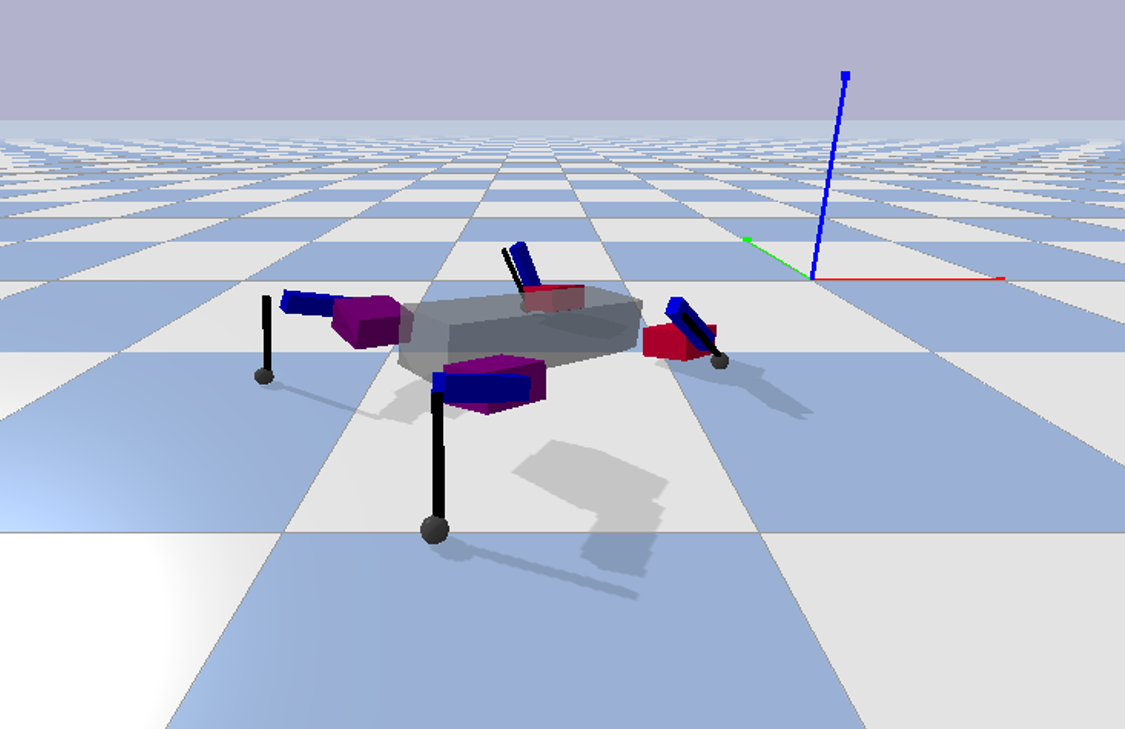}}
\end{picture}
\caption{This model of the Vision 60 quadrupedal robot~\citep{ma2019first}, as rendered in the PyBullet physics engine~\citep{coumans2019}, has been modified to widen the base. This eliminates unrecoverable states in which the robot has tipped over, which was necessary when using uniformly random control inputs as we do to generate a training set of the robot's dynamics.}
\label{fig:quadruped}
\end{figure}

\subsection{Quadrupedal robot}
\label{sec:quadbot}

We consider the problem of predicting the trajectory of the center of mass of a quadrupedal robot with 12 actuators. We have used a modified model of the Vision 60 quadrupedal robot~\citep{ma2019first} simulated in the PyBullet physics engine~\citep{coumans2019}. A screenshot of the Vision 60 model in the PyBullet engine is shown in Fig.~\ref{fig:quadruped}. There is one rotary actuator under velocity control at each knee and each hip and shoulder joint, and one rotary actuator under position control governing abduction and adduction of each hip and shoulder joint. We use 0.075~s timesteps which have been simulated more finely by the physics engine.

Similarly to the double-well potential experiment, we chose uniformly random control inputs and simulated the dynamics for 500,000 timesteps for use in supervised training. After training each model to convergence we compared their respective performance.

We observe, as shown in Fig.~\ref{fig:quadbot-results}, that the \textit{Extended Koopman Model} is once again able to accurately capture the nonlinear, nonconvex dynamics, predicting accurately out to 120 time steps. The \textit{Convex Koopman Model} performs reasonably well, following the general trend and resolving some of the larger changes, but failing to resolve short-term dynamics. The traditional Koopman model performs quite poorly, only coarsely fitting even the first 10 timesteps.

\section{Discussion}

This paper has introduced two novel generalizations of traditional Koopman models, and has shown experimentally that these two generalizations significantly improve predictive performance on two nonlinear, nonconvex dynamic systems. 

These generalizations preserve an unique trait of Koopman models: the potential for fast, globally optimal control. Prior work has demonstrated Koopman-based model predictive control~\citep{proctor2018generalizing,peitz2019koopman,korda2018linear}, but the low predictive performance of traditional Koopman models on even simple nonlinear, nonconvex dynamic systems has limited its potential. This paper presents a solution to the issue of low predictive performance; we have shown experimentally that both \textit{Convex Koopman Models} and \textit{Extended Koopman Models} significantly outperform traditional Koopman models for two nonlinear, nonconvex systems, including a complex electromechanical system.

This paper can also be viewed as generalizing another line of work: the use of input convex neural networks for model predictive control. Recently~\citet{chen2020input} have demonstrated that the dynamics of certain power distribution systems are convex and can be modeled with input convex neural networks. This leads to a fast algorithm for optimal voltage regulation via convex optimization. \textit{Convex Koopman Models} generalize this method to work with nonconvex systems; one lifts the dynamics of a nonconvex system to a space in which its dynamics are approximately convex. \textit{Extended Koopman Models} further generalize this approach, potentially enabling fast model predictive control of highly nonlinear, nonconvex systems.

As future work, we plan to validate this approach's usefulness for model predictive control. Thus far we have demonstrated that these generalizations to Koopman models significantly improve predictive performance for two systems. However, we currently only have theoretical justification that this will lead to practical optimal control of these systems. We expect that getting good performance from a model predictive controller based on an \textit{Extended Koopman Model} may require further innovation; the fact that the control transformation is only approximately invertible could pose unforeseen difficulties. We expect that getting good performance from a model predictive controller based on a \textit{Convex Koopman Model} will be fairly straightforward and will lead to significant improvements to globally optimal control based on input convex neural networks, and to globally optimal control based on Koopman methods.

These are not the only possible generalizations to Koopman models. We hope that this work will prove useful in practice for globally optimal control; we also hope that this work will inspire others working with models with desirable special properties to look for generalizations which still satisfy those properties. 

\section*{Acknowledgements}

The authors would like to thank Fonterra for supporting this research.

\bibliography{mybib}

\end{document}